\documentclass[twoside,fleqn]{article}
\usepackage{espcrc2} 

\newcommand{\eqn}[1]{(\ref{#1})}
\newcommand{\ft}[2]{{\textstyle\frac{#1}{#2}}}
\def\be{\begin{equation}}
\def\ee{\end{equation}}
\def\bea{\begin{eqnarray}}
\def\eea{\end{eqnarray}}

\renewcommand{\a}{\alpha}

\renewcommand{\d}{\delta}
\newcommand{\pa}{\partial}

\newcommand{\e}{\epsilon}

\newcommand{\m}{\mu}
\newcommand{\n}{\nu}

\def\IZ{{\hbox{{\rm Z}\kern-.4em\hbox{\rm Z}}}}

\def\bigone{{\hbox{1\kern -.23em{\rm l}}}}
\title{Electric-Magnetic Dualities in Supergravity }
\author{Bernard de Wit\address{Institute for Theoretical Physics and
Spinoza Institute \\
Utrecht University, Utrecht, The Netherlands }}
\begin{document}
\setlength{\arraycolsep}{0pt}
\begin{titlepage}
\begin{flushright} ITP-UU-01/08 \\
                  SPIN-01/05 \\[1mm]
 hep-th/0103086
\end{flushright}
\vfill
\begin{center}
{\large\bf Electric-Magnetic Dualities in Supergravity${}^\dagger$}\\
\vskip 7.mm
{Bernard de Wit }\\
\vskip 0.1cm                                                      
{\em Institute for Theoretical Physics and Spinoza Institute} \\
{\em Utrecht University, Utrecht, The Netherlands} 
\end{center}
\vfill
\begin{center}
{\bf ABSTRACT}
\end{center}
\begin{quote}
I review electric-magnetic duality from the perspective of extended
supergravity theories in four spacetime dimensions. 
\vfill      \hrule width 5.cm
\vskip 2.mm
{\small\small
\noindent $^\dagger$ Invited talk given at Thirty Years of
Supersymmetry, October 13 - 15, 2000, Minneapolis; to be published  
in the proceedings.}  
\end{quote}
\begin{flushleft}
\today \\
February 2001
\end{flushleft}
\end{titlepage}
\begin{abstract}
I review electric-magnetic duality from the perspective of extended
supergravity theories in four spacetime dimensions. 
\end{abstract}
\maketitle
\setcounter{footnote}{0}
\section{INTRODUCTION}
In early June 1974, Frank Yang returned from the London Conference and
reported during an ITP luncheon meeting in Stony Brook on the two, in his
view, most exciting developments: the 't~Hooft-Polyakov monopole and
supersymmetry. At that point I decided to start and look more closely into the
papers of Wess and Zumino, which had already intrigued me during that
past year. In August I spent some time in Aspen where I was the only
participant working on supersymmetry, although Max Dresden and  Lochlainn
O'Raifeartaigh expressed a clear interest in the subject. In the fall  
of that year I published the results of my first research, which I
believe was the first paper on supersymmetry in 
four spacetime dimensions that appeared in the Physical Review. 
During that academic year I started working with Dan Freedman; this 
marked the beginning of a series of fruitful collaborations. In the
summer of 1975 I returned to Leiden, where I made an excursion with
Gerard 't Hooft into lattice gauge theories. But in view of the  
exciting discoveries made in supersymmetry and supergravity, it did not
take long before I returned to these topics. In 1977 I again attended
the Aspen summer program, where now everybody was working on
supersymmetry. When driving towards Aspen from Stony Brook with Dan  
Freedman we started to discuss the construction of $N\!=\!8$
supergravity. The first person whom we met at the Center in 
Aspen was Murray Gell-Mann, who, after inquiring  
what we were working on, expressed his amazement that we were jumping
right ahead from $N\!=\!4$ (which had barely been constructed
\cite{Das,CSF1}) to $N\!=\!8$ supergravity. We had  
many discussions with the participants about our work in which we
derived the $N\!=\!8$ supergravity Lagrangian and transformation rules to
second order using the Noether method \cite{DWF}. Our starting point
was the observation that the scalar and pseudoscalar fields had to
transform under SO(8) as self-dual four-rank antisymmetric tensors. 
We did not know what the
self-duality phases were for the scalar and pseudoscalar fields and
initially assumed that they were equal. However, then we got stuck; it
was Martin Ro\v{c}ek  who suggested 
to us that perhaps the phases had to be opposite, at which point it
occurred to us that opposite phases were indeed crucial for the
existence of SU(8) electric-magnetic duality. Electric-magnetic
duality had already been established at that time for $N\leq4$
\cite{FSZ,CSF1}.\footnote{ 
  The first time that electric-magnetic duality appeared in 
  supergravity was probably in the study of the one-loop divergences
  \cite{vNV} of $N\!=\!1$ Maxwell-Einstein supergravity and the one-loop
  finiteness \cite{GvNV} of pure $N\!=\!2$ supergravity \cite{FvN}. } 

This was my first encounter with electric-magnetic duality, which
would turn out to be such an important feature of extended supersymmetric
field theories. When preparing my presentation for this symposium I
was struck by just how many papers have dealt with this subject
since that time. Clearly it is hard to do justice to all this
work. I stress that this modest effort represents very much my
own personal perspective, but I made an effort to collect many of the
relevant references at the end. 

After these reminiscences let me turn to electric-magnetic
duality. It is well known that in Maxwell theory in four-dimensional
(flat or curved) Minkowski space, one can perform 
(Hodge)  duality rotations, which commute with the Lorentz group and
rotate the electric and magnetic fields and inductions according to
\be
{\bf E} \,\longleftrightarrow\, {\bf H} \quad ,\quad 
{\bf B} \longleftrightarrow {\bf D} \quad.
\ee
In the following I will discuss generalizations of this Maxwell
duality. 

Throughout this paper I consider field theories with abelian vector
gauge fields labeled by indices $I,J,\ldots$. All fields are neutral,
{\it i.e.} the gauge fields enter the Lagrangian only through their
(abelian) field strengths and not through covariant derivatives. These
field strengths $F_{\m\n}$ are decomposed into selfdual and
anti-selfdual components (which are related by complex conjugation)
and so are the field strengths $G_{\m\n}$ 
that appear in the field equations (together 
$F_{\m\n}$ and $G_{\m\n}$ comprise the electric and magnetic fields
and inductions), 
\bea
F_{\m\n} &\longrightarrow\,&  F_{\m\n}^{+I} + F_{\m\n}^{-I}\;,
\nonumber\\[1mm] 
G_{\m\n} &\longrightarrow\,&  \left\{
\begin{array}{rcl}
G_{\m\n \,I}^+ &=& 2 i \displaystyle{\pa L\over \pa F^{+\m\n\,I}
}\;,\\[2mm] 
G_{\m\n\,I}^- &=& -2 i \displaystyle{\pa L\over \pa F^{-\m\n \,I} }\;. 
\end{array} \right.
\eea
The Bianchi identities and equations of motion
for the abelian gauge fields take the form
\begin{equation}
\partial^\mu \big(F^{+} -F^{-}\big){}^I_{\m\n} =
\partial^\mu \big(G^+ -G^-\big){}_{\m\n I} =0\,.
\label{Maxwell}
\end{equation}
The above equations are obviously invariant under
the following rotations of the field strengths, 
\bea
\pmatrix{F^{\pm I}_{\mu\nu}\cr \noalign{\vskip 1mm} G^\pm_{\mu\nu I}\cr} 
\longrightarrow \pmatrix{\tilde F^{\pm I}_{\mu\nu}\cr \noalign{\vskip
1mm} \tilde G^\pm_{\mu\nu I}\cr} = 
\pmatrix{U&Z\cr \noalign{\vskip 1mm} W&V\cr} \pmatrix{F^{\pm  
I}_{\mu\nu}\cr  \noalign{\vskip 1mm} G^\pm_{\mu\nu I}\cr}\,, \nonumber \\
\label{FGdual}
\eea
where $U^I_{\,J}$, $V_I^{\,J}$, $W_{IJ}$ and $Z^{IJ}$ are 
constant real  $n\times n$ submatrices and $n$ denotes the number of
independent gauge potentials.\footnote{
  Note that I use the tilde to denote the new rotated field strengths,
  and {\it not} the Hodge-dual of the field strengths! This notation
  is used throughout. Furthermore, observe that electric-magnetic
  duality can be formulated straightforwardly in curved space; it is
  simply a matter of convenience  that I am assuming flat space
  here. For an early discussion of electric-magnetic duality in curved
  space, the reader may consult \cite{MisnerWheeler}. } 
The question is, however, whether the rotated equations can again
follow from a Lagrangian. More precisely, does there exist a 
Lagrangian $\tilde L(\tilde F)$ depending on the new field strengths
$\tilde F^I_{\m\n}$ such that the new tensors $\tilde G_{\m\n I}$
follow from $\tilde L$ 
by differentiation with respect to $\tilde F^{\m\n\, I}$, {\it i.e.}, 
\be 
\tilde G_{\m\n\,I} \propto {\pa \tilde L(\tilde F)\over \pa \tilde
F^{\m\n\,I}}\,. 
\ee
This condition amounts to an integrability condition, because one can
derive an expression for the derivative of $\tilde L$ with respect to the
original field strength $F^I_{\m\n}$, which can only have a solution
(for nontrivial Lagrangians) provided that the matrix in \eqn{FGdual}
is an element of the group 
${\rm Sp}(2n;{\bf R})$.\footnote{
  I exclude transformations with $U=V\propto {\bf 1}$ and
  $W=Z=0$, since this amounts to a simple
  rescaling of the Lagrangian and field strengths. } 
The symplectic group implies the following condition on the submatrices,
\bea
&&U^{\rm T} V -W^{\rm T}Z = VU^{\rm T} - WZ^{\rm T} = {\bf 1}\,,
\nonumber\\  
&&U^{\rm T}W = W^{\rm T}U \;,\qquad Z^{\rm T}V =V^{\rm T}Z \,.
\eea
The fact that the symplectic redefinitions of the field strengths
constitute the group ${\rm Sp}(2n;{\bf R})$ was first derived in
\cite{GZ}, but in the context of a duality {\it invariance} rather
than of a {\it reparametrization}. In this respect our presentation
is more in the spirit of the later treatment in \cite{CecFerGir} for
$N=2$ vector multiplets coupled to supergravity \cite{DWVP}. Given
that the duality belongs to the class of symplectic
reparametrizations, one can derive that the new Lagrangian is of the
form,  
\bea
&&\tilde L(\tilde F) + \ft14 i (\tilde F^{+I}_{\m\n} \tilde G^{+\m\n}_I
- \tilde F^{-I}_{\m\n} \tilde G^{-\m\n}_I) \nonumber \\[.5mm]
&& = L(F) +\ft14 i(F^{+I}_{\m\n} G^{+\m\n}_I - F^{-I}_{\m\n}
G^{-\m\n}_I )\,,   \label{newlagrangian}
\eea
up to terms independent of $F^I_{\m\n}$. The equation 
\eqn{newlagrangian} is  analogous to a result known for $N\!=\!2$
vector multiplet Lagrangians \cite{dW96}; this will be discussed in
section~4. 

The above expression \eqn{newlagrangian} is not so useful, as it 
requires substituting $\tilde F^I_{\m\n}$ in terms of $F^I_{\m\n}$, or
vice  versa. Furthermore it is clear that the Lagrangian does not transform
as a function, since 
\be
\tilde L(\tilde F) \not= L(F)\,.
\ee
However, the linear combination of the Lagrangian and the
$F^I_{\m\n}\, G^{\m\n}_I$ terms, 
\be 
L(F) +\ft14 i(F^{+I}_{\m\n} G^{+\m\n}_I - F^{-I}_{\m\n}
G^{-\m\n}_I )\,, \label{FG-function}
\ee
does transform as a function. 

When $L$ remains unchanged, {\it i.e.} when 
\be
\tilde L(\tilde F) = L(\tilde F)\,, \label{dual-invariant}
\ee
then the  theory is {\it invariant} under the 
corresponding transformations. Again this is hard to verify 
explicitly in this general form. A more convenient method instead, is
to verify  that the substitution $F^ I_{\m\n}\to \tilde F^I_{\m\n}$
into the derivatives  
$\pa L(F)/\pa F^{\m\n I}$ correctly induces the symplectic
transformations of the field strengths $G_{\m\n I}$. Furthermore, the
linear combination \eqn{FG-function}  
must be an {\it invariant function} under $F_{\m\n}^I \to \tilde
F_{\m\n}^I$. Note that in the literature the word duality is used both
for equivalence and for invariance relations. Whenever the context is
not clear I will always try and indicate specifically whether I am 
dealing with an equivalence or an invariance. 

Observe that the presence of other fields does not play an immediate
role here. Their symmetry transformations must be combined with the
duality transformations. The transformations 
of the other fields have been incorporated already in the duality
transformations \eqn{FGdual}, as the field strengths $G_{\m\n I}$ may depend
on other fields. In that case the variations of these other fields are
crucial in that they must generate (together with the transformation of
the field strengths $F^I_{\m\n}$) the correct transformation rules for the
tensors $G_{\m\n I}$. Once this is accomplished, the result
\eqn{newlagrangian} will 
still apply and the $F^I_{\m\n}$-independent terms that
are not constrained by \eqn{newlagrangian} must be separately
invariant under the transformations acting on the fields other than
the field strengths. I return to this issue in the next section,
where I will be be dealing with a more specific set of Lagrangians.  

When the duality invariance is continuous, one may evaluate the effect
of an infinitesimal transformation. Hence one expands, 
\be
\pmatrix{U&Z\cr \noalign{\vskip 1mm} W&V\cr} \approx {\bf 1} +
\pmatrix{B&-D\cr \noalign{\vskip 1mm} C&-B^{\rm T}} \,, \label{infi}
\ee
where the matrices $C_{IJ}$ and $D^{IJ}$ are symmetric. Invariance
requires now that the following linear combination vanishes,  
\bea
&&C_{IJ} (F^{+I}_{\m\n} F^{+J\m\n}-F^{-I}_{\m\n} F^{-J\m\n}) \nonumber \\
&& -2 B^I_{\,J}( G^{+}_{\m\n I} F^{+J\m\n} - G^{-}_{\m\n I}
F^{-J\m\n}) \nonumber \\
&&+ D^{IJ}( G^{+}_{\m\n I} G^{+\m\n}_J - G^{-}_{\m\n I} G^{-\m\n}_J )
= 0\,. \label{consistency}
\eea

In case the theory has electrically and/or magnetically charged
states, these would show up as source terms in the Maxwell
equations. Consequently, these magnetic and electric charges, which I
denote by $(p^I,q_I)$, should
rotate under the action of the duality group. However, it is well
known that these charges have to satisfy Dirac-Schwinger-Zwanziger
quantization conditions and thus span some lattice. This lattice is then
left invariant by an arithmetic subgroup ${\rm Sp}(2n;{\bf Z})$ of the
duality group. 

A relevant example of a Lagrangian that depends exclusively on a field
strength, is the Born-Infeld Lagrangian of nonlinear
electrodynamics, defined by \cite{BI}, 
\be 
L =- g^{-2} \sqrt{\det[\eta_{\m\n} + g\, F_{\m\n}]} + g^{-2}\,,
\label{BI-L}
\ee
where $g$ is a coupling constant. 
It has been known for a long time that this Lagrangian has
electric-magnetic duality symmetries \cite{schroedinger} (for more
recent references, see \cite{GR,GZ2}). To deal with this aspect let us
first rewrite the determinant as follows, 
\bea
\Delta&=&\det[\eta_{\m\n}+ g\,F_{\m\n}] \nonumber\\
&=& 1 -\ft12 g^2 {\rm Tr} (F^2) 
+\ft18 g^4 [{\rm Tr} (F^2) ]^2  -\ft14 g^4{\rm Tr} (F^4) \nonumber \\
&=&  1 + \ft12 g^2(F^{+2} + F^{-2}) + \ft1{16}g^4 (F^{+2} - F^{-2})^2
\,,\nonumber\\
&~&
\eea
where $F^{\pm 2} = F^\pm_{\m\n} F^{\pm \m\n}$. 
The expression for $G_{\m\n}$ follows then directly, 
\be
G^+_{\m\n}= -  {i \over \Delta^{1/2}} F^+_{\m\n} \Big[1 + \ft14 g^2
(F^{+2} - F^{-2}) \Big]\,.
\ee
Inserting this expression into the condition \eqn{consistency} it
follows that there is only one continuous duality symmetry
characterized by $C=D$ and $B=0$. Obviously, this leads to finite
transformations corresponding to ${\rm SO}(2)$. There exist no
new discrete duality symmetries. 

Under general electric-magnetic duality transformations corresponding
to ${\rm Sp}(2;{\bf R}) \cong {\rm SL}(2,{\bf R})$, one obtains a variety
of different but inequivalent Lagrangians whose parameter space is
isomorphic to ${\rm SL}(2)/{\rm SO}(2)$. However, it is rather
cumbersome to work out this parametrization explicitly.

\section{U$(N)$ DUALITY} 
The Lagrangians of $N$-extended supergravity in four spacetime
dimensions contain $\ft12N(N-1)$ abelian vector gauge
fields\footnote{
  For $N\!=\!6$ the number is different and equal to 16; I disregard 
  $\!N=\!7$ which is just a relabeling of the $N\!=\!8$
  theory.} 
and depend at most quadratically on the field
strengths. I parametrize them according to 
\bea
{L}&=& -\ft14 i\sqrt{g} \, {\cal N}_{IJ}\,F_{\mu\nu}^{+I}F^{+\mu\nu
  J}  -\ft12 i\sqrt{g}\, F_{\mu\nu}^{+I} {\cal O}^{+\mu\nu}_I  \nonumber \\
&& + \mbox{ h.c.} + F\mbox{-independent terms}\,, \label{FF+FO}
\eea
where the $F_{\m\n}^{\pm I}$ are the (anti-)selfdual field strengths  
and $\cal N$ may depend on the scalar fields. In addition
there are moment couplings  
({\it e.g.} to the fermions) encoded in tensors ${\cal
O}^{\pm\m\n}_I$ whose form is left unspecified.  
For these Lagrangians one can investigate whether they are invariant
under a certain class of electric-magnetic duality
transformations. One obvious invariance group is 
${\rm SO}(N)$ with the vector potentials transforming in the
adjoint representation. For supergravity this group can be extended to 
${\rm U}(N)$ which, in view of the reality of the vector potentials,
cannot act on the vector potentials but only on the field strengths
by means of electric-magnetic duality. Hence the invariance
does not apply directly to the Lagrangian, but to 
the combined equations of motion and Bianchi identities. 

There is a number of reasons to expect that ${\rm U}(N)$ is in fact
the minimal invariance group of the field equations. First of all,
${\rm U}(N)$ is the subgroup of the automorphism group of the
supersymmetry algebra, which acts chirally on the $N$ supercharges and
commutes with the Lorentz group.  Therefore massless supermultiplets
consist of states that are assigned to representations of ${\rm U}(N)$
\cite{HaagLS}. This assignment can then be extended to the underlying
fields (assuming one can find suitable field representations). It is
important to realize that this by itself does not imply that the
action or the equations of motion are invariant under ${\rm U}(N)$,
because the theory could allow for certain deformations that break
the invariance (the most obvious deformation is a gauging of 
${\rm SO}(N)$ associated with the vector fields). However, the
simplest such theory will exhibit ${\rm U}(N)$ as an invariance group. 
Secondly, ${\rm U}(N)$ is a local gauge invariance of the
superconformal theories, which, at least for $N\leq4$, govern the
leading spin components of the Poincar\'e supergravity theories
\cite{FKTvN,BDRDW}. 

Indeed $N$-extended supergravity has been shown to exhibit ${\rm U}(N)$ 
duality for $N=2,3$ in \cite{FSZ} and for $N=4$ in \cite{CSF1}. For
$N=8$ the situation was analyzed in \cite{dW79}. To be more specific,
let me return to the Lagrangians \eqn{FF+FO}. Obviously, the
corresponding tensors $G_I$ take the form,  
\bea
G^+_{\mu\nu I}&=&{\cal N}_{IJ}F^{+J}_{\mu\nu} + {\cal O}_{\mu\nu 
I}^+\,, \nonumber \\
G^-_{\mu\nu I}&=&\bar {\cal N}_{IJ} F^{-J}_{\mu\nu} 
+ {\cal O}_{\mu\nu I}^- \,. \label{defG}
\eea
The Bianchi identities and equations of motion
for the abelian gauge fields are given in \eqn{Maxwell}.  

{From} (\ref{defG}) and (\ref{FGdual}) one derives 
that $\cal N$ must transform according to a fractional linear
transformation, 
\begin{equation}
\tilde{\cal N}_{IJ} = (V_I{}^K {\cal N}_{KL}+ W_{IL} )\,
\big[(U+ Z {\cal N})^{-1}\big]^L{}_{\!J}  \,,\label{nchange}
\end{equation}
and this transformation must be induced by proper changes of
the scalar fields. 
To ensure that $\cal N$ remains a symmetric tensor, at 
least in the general case, the transformation  
(\ref{FGdual}) must be an element of ${\rm Sp}(2n;{\bf R})$,  in accord
with what was claimed in the previous section  
(again I disregard uniform scale transformations). 
Furthermore the tensor $\cal O$ must change according to
\be
\tilde{\cal O}_{\mu\nu I}^+ = {\cal O}_{\mu\nu J}^+  \,[(U+Z{\cal 
N})^{-1}]^J{}_{\!I}\,, \label{ochange}
\ee
and likewise for ${\cal O}^-$. The function $\cal N$ defines
(possibly field-dependent) expressions for the generalized coupling
constants and $\theta$-angles,  ${\cal N}_{IJ} \propto
\theta_{IJ}/2\pi+i4\pi/g_{IJ}^2$. The sign of the coupling constants
does not change under electric-magnetic duality, as follows from   
\bea
&&(\tilde{\cal N}-\tilde{\bar{\cal N}})_{IJ} = ({\cal
N}-{\bar{\cal N}})_{LK}  \nonumber \\
&&\qquad \times\big[(U+ Z{\cal N})^{-1}\big]^L{}_{\!I} \,\big[(U+
Z \bar{\cal N})^{-1}\big]^K{}_{\!J}  \,.\label{gchange}  
\eea

It is possible to rewrite the Lagragian \eqn{FF+FO} as follows, 
\bea
L &=& -\ft14 i\sqrt{g}  F^{+I}_{\m\n} G^{+\m\n}_I - \ft14i \sqrt{g} 
F^{+I}_{\m\n} O^{+\m\n}_I + \mbox{h.c.} \nonumber\\[1mm]
&=& -\ft14i \sqrt{g} ( F^{+}_{\m\n} + F^{-}_{\m\n})^I ( G^{+\m\n} -
G^{-\m\n})_I \nonumber \\
&& -\ft14\sqrt{g}\Big[ i O^+_{\m\n I}[({\cal N}-\bar {\cal
N})^{-1} ]^{IJ} 
\nonumber\\ 
&&\hspace{12mm} \times  ( G^{+\m\n}_J 
-\bar {\cal N}_{JK} F^{+\m\n K})  + \mbox{h.c.} \Big] \,,
\label{FG-Lagr} 
\eea
where in the second equation I introduced an $F^I_{\m\n}$-independent
term by hand so that the last term  transforms as
a scalar under symplectic reparametrizations. The first term in the 
last equation is {\it not} a scalar, but it is equal to a
total derivative by virtue of Maxwell's equations. Any other
$F^I_{\m\n}$-independent terms in 
the Lagrangian \eqn{FF+FO} must separately transform as a scalar.

To ensure invariance under ${\rm U}(N)$ duality transformations one
must first verify whether the transformations of the scalars induce
the correct variation \eqn{nchange} of ${\cal N}_{IJ}$.
Secondly, one must verify that all $F^I_{\m\n}$-independent terms other than
the ones already included in the last term in \eqn{FG-Lagr}, are
separately invariant 
under ${\rm U}(N)$ transformations of the remaining fields. 
For Lagrangians which depend at most quadratically on the field
strengths (such as the one discussed in this section), infinitesimal
duality symmetries must satisfy the condition (cf. \eqn{consistency}), 
\bea
&&C_{IJ} \,F^{+I}_{\m\n} F^{+J\m\n} -2 B^I_{\,J}\, G^{+}_{\m\n I}
F^{+J\m\n}  \nonumber \\
&&+ D^{IJ}\, G^{+}_{\m\n I} G^{+\m\n}_J = 0\,. \label{consistency2}
\eea
This condition is the direct analogue of the so-called consistency condition
that I will present in section~4 (cf. \eqn{consistency3}) in the
context of $N\!=\!2$ vector multiplets coupled to supergravity
\cite{DWVP}. This condition was explored extensively in \cite{bigE}.   

It is obvious that electric-magnetic duality invariances can be
broken. This is rather straightforward for low-$N$ supergravity where
one can couple to supersymmetric matter. But in all cases one can
introduce nonabelian interactions for the gauge fields so 
that the vector potentials (which are real and cannot transform under
the full duality group) appear explicitly in the field equations and
Bianchi identities. Another question is, however, whether the 
electric-magnetic dualities can constitute a group {\it bigger} than
${\rm U}(N)$. With hindsight it is obvious that a bigger group is in
fact natural. For $N\geq4$ the pure supergravity theories contain 
scalar fields which (see Table~1) transform under ${\rm U}(N)$. The
Lagrangian depends on the  scalar fields in a nonlinear
fashion and their kinetic terms take the form of a nonlinear sigma
model with the scalars parametrizing some target 
space. In view of the fact that the Lagrangians become more and more
restricted for large $N$, one expects the target space to exhibit a
high degree of symmetry. Specifically, one 
expects the target spaces to be homogeneous, which means that every
two points of the space can be related by a symmetry transformation
({\it i.e.}, an isometry). 
Because the ${\rm U}(N)$ transformations
act linearly on the scalar fields they only constitute a subgroup of
the full isometry group of the scalar manifold. A homogeneous space
has also nonlinearly realized symmetries, so that its symmetry group is
bigger than ${\rm U}(N)$. As the ${\rm U}(N)$ subgroup is
realized by means of electric-magnetic duality, also the nonlinearly
realized symmetries of the scalar manifold are contained in the
duality group.  

\section{HIDDEN SYMMETRIES}
As argued above pure supergravity theories with $N\geq4$ have
scalar fields that parametrize a homogeneous manifold. The symmetries
of these manifolds include the group ${\rm U}(N)$. The structure of
the scalar sector of the theory is thus based on a coset space G/H,
where G is the ``hidden symmetry'' group and ${\rm H}\!=\! {\rm U}(N)$,
because ${\rm U}(N)$ is the group 
that can act on the other fields (in particular on the fermions by
means of chiral transformations), as I discussed previously. 

\begin{table*}[htb]
\caption{The scalar fields for $N\!=\!4$, 5, 6, 8. The case $N\!=\!7$
coincides with $N\!=\!8$. }
\label{table:1}
\newcommand{\mm}{\hphantom{$-$}}
\newcommand{\cc}[1]{\multicolumn{1}{c}{#1}}
\renewcommand{\tabcolsep}{2pc} 
\renewcommand{\arraystretch}{1.2} 
\begin{tabular}{@{}llcl}
\hline
$N$           & \cc{scalar field} & \cc{U$(N)$ decomposition} &
\cc{homogeneous space}  \\
\hline
4      & \mm $\phi$ & \mm ${\bf 1} + \overline{\bf 1}$ & \mm ${\rm
  SU}(1,1)/{\rm U}(1)$ \\ 
5      & \mm $\phi^i$ & \mm ${\bf 5} + \overline{\bf 5}$ & \mm ${\rm
  SU}(5,1)/{\rm U}(5)$ \\ 
6      & \mm $\phi^{ij}$ & \mm ${\bf 15} + \overline{\bf 15}$ & \mm ${\rm
  SO}^\ast(12)/{\rm U}(6)$ \\ 
8      & \mm $\phi^{ijkl}$ & \mm ${\bf 35} + \overline{\bf 35}$ & \mm ${\rm
  E}_{7(7)} /{\rm SU}(8)$ \\
\hline
\end{tabular}\\[2pt]
\end{table*}

These ``hidden'' symmetries were first discovered for $N\!=\!4$, where
the complex scalar parametrizes the ${\rm SU}(1,1)/{\rm U}(1)$
coset space \cite{CSF2}. The absence of the ${\rm SU}(4)$ group is of
no concern here, because this group can be included as a factor in
both G and H. Hence ${\rm G}\!=\!{\rm SU}(1,1)\times {\rm
SU}(4)$. Interestingly enough, it turned out later that  
the superconformal $N=4$ theory also contains a complex scalar which
parametrizes the same coset space \cite{BDRDW}. Subsequently it was
established that the same situation arises for the complex scalar in
10-dimensional IIB supergravity \cite{SchwarzWest}. 

{From} the above arguments it follows that the group G must satisfy the
following embedding condition,
\be
{\rm U}(N)  \subset {\rm G} \subset {\rm Sp}(2n;{\bf R})
\ee
where, for $N=8$, $n \!=\!28$. The major advance came with the work of
\cite{CJ}, where it was realized that for $N=8$ one has ${\rm G} =
{\rm E}_{7(7)}$, which is indeed a subgroup of ${\rm Sp}(56;{\bf R})$
and contains ${\rm H}= {\rm SU}(8)$ as its maximal compact
subgroup. The fact that the ${\rm U}(1)$ subgroup of ${\rm U}(8)$ is
absent may seem    
surprising in view of the ${\rm U}(8)$ symmetry group 
of the $N=8$ supergravity multiplet, but one must realize that, for
this multiplet,  the ${\rm U}(1)$ subgroup coincides with the ${\rm
SO}(2)$ helicity group. The dimension of ${\rm E}_{7(7)}$ and ${\rm
SU}(8)$ is equal to 133 and 63, respectively, which shows that the
coset space parametrized by the scalar fields is of dimension 70. 

It is often convenient to employ the conjectured duality invariance in 
constructing the full supergravity Lagrangians, as the invariance
gives a handle on the nonlinear interactions with the scalar
fields. There are 
many examples in the literature where this approach was followed. With
regard to $N\!=\!8$ supergravity, the first iterative results for the
Lagrangian and the transformation rules appeared in \cite{DWF}. 
The nonlinear terms were uncovered in \cite{CJ}, where the theory was
constructed by dimensional reduction from eleven dimensional
supergravity \cite{CJS}. In the pioneering paper \cite{CJ}
many of the details of the dimensional reduction of supergravity were
worked out for the first time. The ${\rm E}_{7(7)}$ duality was
conjectured based on a counting argument indicating that the
dimension of the duality group should be 133. This was then confirmed
by working out certain terms of the Lagrangian and of the supersymmetry
transformation rules. The ${\rm SU}(8)$ duality of the theory was
studied in \cite{dW79}. Finally complete result based on ${\rm
E}_{7(7)}$ duality were presented in \cite{dewnic3} and served as a
starting point for the ${\rm SO}(8)$ gauging of $N=8$
supergravity. The resulting theory 
was written in a form where ${\rm SO}(8)\times{\rm SU}(8)$ was
manifest and the ${\rm SO}(8)$ gauge group was embedded into ${\rm
E}_{7(7)}$, thus breaking the electric-magnetic duality to ${\rm
SO}(8)$.  

{From} a more recent perspective the question arises what the
implications of electric-magnetic duality are for eleven-dimensional
supergravity and/or M-theory. In this connection I note that
there exists an alternative formulation of 11-dimensional
supergravity, which does not truncate the theory to the massless
sector in a compactification to four spacetime dimensions while
still preserving certain features of the electric-magnetic 
duality group \cite{dewnic1}. It is based on replacing the ${\rm
SO}(1,10)$ tangent-space symmetry of the eleven-dimensional theory
according to  
\bea
{\rm SO}(1,10) &\longrightarrow\,& {\rm SO}(1,3)\times {\rm SO}(7)
\nonumber\\
&\longrightarrow\,&   {\rm SO}(1,3)\times {\rm SU}(8)\,.
\eea
The construction thus stresses a $4\!+\!7$ split of the $d\!=\!11$
coordinates which leads to corresponding decompositions of the
tensors and spinors. Other splits have been worked out along similar
lines \cite{nic1}. 
The alternative theory is gauge equivalent to the original one
of \cite{CJS} and the equivalence holds at the level of the equations
of motion. The 
reformulation can be regarded as a first step towards fusing the
bosonic degrees of freedom of $d=11$ supergravity in a way which is
more in harmony with the hidden symmetries of the dimensionally
reduced theories. 

An important quantity is the so-called generalized
vielbein, denoted by $e^m_{AB}$, which replaces the internal
sevenbein. It carries an upper world index $m=
1,\ldots,7$ and lower ${\rm SU}(8)$ indices $A,B=1,\ldots,8$. It is
antisymmetric in the indices $A,B$ so that it transforms in the $\bf
28$ representation of ${\rm SU}(8)$, 
\be
e^m_{AB}\rightarrow U_A{}^C \,U_B{}^D \,e^m_{CD}\,,
\ee
where $U_A{}^C$ is an SU(8) matrix depending on all eleven 
coordinates. 
By including its complex conjugate,
\be
e^{mAB} = (e^m_{AB})^*\,,
\ee
the generalized vielbein is thus given by
the pseudo-real tensor $(e^{mAB} , \,e^m_{AB})$, which, for given $m$,
constitutes the ${\bf 56}$ 
(pseudo-real) representation of ${\rm E}_{7(7)}$.

The generalized vielbein has many more components than the original 
siebenbein, but of course the number of physical degrees of freedom
is the same as before. Some of the redundant degrees of freedom
are taken care of by the SU(8) gauge symmetry, and in addition the
generalized vielbein is subject to a number of constraints. One is the
Clifford property, already identified in \cite {dewnic1}:
\be
e^m_{AC}\, e^{nCB} + e^n_{AC}\, e^{mCB} = \ft14 \delta_A^B
\,e^m_{CD}\, e^{nDC}\,.  
\ee
One can also derive a formula for the
original seven-metric, 
\be
({\rm det} \, g)^{-1/2}\, g^{mn} = \ft18 \,  e^m_{CD} \,e^{nCD} \,,
\ee
which can serve as a master formula for the full non-linear metric ansatz   
in supersymmetric Kaluzu-Klein compactifications of $d\!=\!11$ supergravity. 
In fact the Clifford property is part of the following set of ${\rm
E}_{7(7)}$ covariant constraints, 
\bea
e^m_{AB} e^{nAB} -e^n_{AB} e^{mAB} &=& 0 \,,\nonumber \\
e^m_{AC} e^{nCB} + e^n_{AC} e^{mCB} - 
     \ft14 \delta_A^B e^{m}_{CD} e^{nDC} &=& 0 \,, \nonumber\\
e^m_{[AB} e^n_{CD]} - \ft1{24} \varepsilon_{ABCDEFGH}
   e^{mEF} e^{nGH} &=& 0\,. 
\eea
These equations correspond to the singlet and the $\bf 133$ 
in the ${\rm E}_{7(7)}$ decomposition, 
\be 
{\bf 56} \otimes  {\bf 56}
\rightarrow {\bf 1} \oplus{\bf 133} \oplus{\bf 1463} \oplus{\bf 1539}\;.
\ee
The constraints can thus be rephrased as the statement that
the product $e^m \otimes  e^n$ only contains the $\bf 1463$ and
$\bf 1539$ representations of ${\rm E}_{7(7)}$. 

In addition to the algebraic constraints, the generalized vielbein
satisfies a set of first-order differential relations, called the
``generalized  vielbein postulate'' in \cite{dewnic1}. In order to state them, 
we need suitable ${\rm E}_{7(7)}$ connections ${Q_M}^{\!A}{}_{\!B}$ and 
$P_M^{ABCD}$ in {\em eleven} dimensions. These are built out of the SO(1,10)
coefficients of anholonomity and the four-index field strength 
$F_{MNPQ}$ of $d\!=\!11$ supergravity in the way explained in \cite{dewnic1}; 
since the explicit expressions are somewhat cumbersome I refer readers
there for details. The vector ${Q_M}^{\!A}{}_{\!B}$ acts as the
connection for the local SU(8) transformations and is therefore in the
$\bf 63$ representation of that group. The tensor $P_M^{ABCD}$
transforms in the $\bf 70$ representation under the action of
SU(8). Together they 
constitute the (adjoint) $\bf 133$ representation of ${\rm E}_{7(7)}$. 
For the massless theory these quantities are directly
related to the pull-backs to $d\!=\!11$ spacetime of the tangent-space
connection and vielbein associated with the homogeneous space 
${\rm E}_{7(7)}/{\rm SU}(8)$. 
 
The generalized vielbein postulate takes the form 
\bea
{\cal D}_\mu e^m_{AB} + \ft12 {\cal D}_n {B_\mu}^n  e^m_{AB} + {\cal
D}_n {B_\mu}^m  e^n_{AB}&& \nonumber \\
+ 2\, Q_{\mu[A}{}^{\!C} e^m_{B]C} + P_{\mu ABCD}\, e^{mCD} &=& 0\,,
\nonumber \\
{\cal D}_n e^m_{AB}
+ 2\, Q_{n[A}{}^{\!C} e^m_{B]C} + P_{n ABCD}\, e^{mCD} &=& 0\,, 
\eea
where $B_\m{}^m$ are the Kaluza-Klein vector fields that originate from
the $d\!=\!11$ metric; furthermore 
\be\label{Dmu}
{\cal D}_\mu = \partial_\mu - {B_\mu}^m {\cal D}_m\,,
\ee
for $\mu=0,1,2,3$ and
\bea 
{\cal D}_m e^n_{AB} &=& \partial_m e^n_{AB} + {\Gamma_{mp}}^n e^p_{AB}
                  + \ft12 {\Gamma_{mp}}^p  e^n_{AB}\,,\nonumber\\
{\cal D}_m {B_\mu}^n &=& \partial_m  {B_\mu}^n + {\Gamma_{mp}}^n
{B_\mu}^p\,, 
\eea 
for the internal indices $m,n,\ldots \!=\!1,\ldots,7$. The extra term
with ${\Gamma_{mp}}^p$ in 
the above relation arises because the generalized vielbein 
transforms as a density. Observe
that the affine connection ${{\Gamma}_{mn}}^p$ still depends on
all eleven coordinates, but is still arbitrary at this point, as it
cancels between the different terms in the generalized vielbein postulate.
A convenient choice is the standard Christoffel connection.

So I conclude that all the quantities introduced above comprise ${\rm
E}_{7(7)}$ representations. I stress once more that I am still dealing
with the full $d\!=\!11$ supergravity theory. This pattern continues. For
instance, the supersymmetry variation of the generalized vielbein
takes a form that closely resembles the four-dimensional
transformation rule for the massless modes (in the truncation to the
massless modes, $e^m_{AB}$ is proportional to the ${\rm
E}_{7(7)}/{\rm SU}(8)$ coset representative),
\be
\delta e^m_{AB} =  - \sqrt{2}\,\Sigma_{ABCD} \, e^{m\,CD}\,,
\ee
where
\be
\Sigma_{ABCD}=\bar \e_{[A} \chi_{BCD]} + \ft1{24} \,\e^E\chi^{FGH}\,,
\ee
where $\e_A$ and $\chi_{ABC}$ denote the supersymmetry parameters and
the spin-1/2 fields, respectively. Similarly, the bosonic and fermionic 
equations of motion can be cast into a fully SU(8) covariant form.

In spite of the fact that the theory can be formulated elegantly in
terms of ${\rm E}_{7(7)}$ quantities, it cannot be invariant under 
${\rm E}_{7(7)}$. An obvious obstacle to the invariance seems to be the
presence of the seven Kaluza-Klein vector fields $B_\m{}^{\!m}$. For
instance, when restricting ourselves to the massless 
modes in the toroidal compactification, which is ${\rm E}_{7(7)}$
invariant, the explicit coupling to these vector fields disappears in
the generalized vielbein 
postulate, and evidently many of the formulas displayed above are
trivially satisfied.\footnote{
I remind the reader that there are 21 additional massless vector
fields in the 
toroidal reduction, which originate from the $d\!=\!11$ tensor
field. However, those fields do not appear explicitly in the
generalized vielbein postulate, and are contained in the ${\rm
E}_{7(7)}$ connections.} 
Retaining the massive Kaluza-Klein states associated
with the torus, one observes that it is not possible to preserve ${\rm
E}_{7(7)}$, because the Kaluza-Klein charges do not cover a complete  
${\rm E}_{7(7)}({\bf Z})$ invariant lattice (as they should in
M-theory). In other words, neither the gauge fields nor the central
charges constitute a representation of the duality group.  Observe
that the momentum lattice of the Kaluza-Klein states would in any case
restrict the duality group ${\rm E}_{7(7)}$ to an arithmetic subgroup. 

An intriguing question is whether one can
somehow extend the Kaluza-Klein states to a fully duality-invariant
set of states. This extension to a BPS-extended supergravity
theory, defined as a duality invariant completion of toroidally
compactified supergravity with BPS states \cite{strings00,istanbul}, will
incorporate more essential 
features of M-theory. Because I did not insist on truncating to the
massless modes in the above setting, it should in principle be possible to
incorporate these extra states and consider other than toroidal
backgrounds. Of course, one should not only extend the configuration
of BPS states, 
but also the gauge fields that can couple to the corresponding central
charges. While this can be done for higher dimensions, in four
dimensions this is difficult because the 56 central charges are related to
electric and magnetic charges and those cannot simultaneously be
realized in a local field theory. Nevertheless, assuming that this is
a way to go, then one should also find traces of the 
duality group in compactifications of $d\!=\!11$ supergravity
on non-trivial internal manifolds. In that respect the above results
are again promising because there no reference was made to any particular
background. To date there is only one nontrivial 
compactification, namely 
the $AdS_4\times S^7$ compactification of $d=11$ supergravity
\cite{dewnic2}  (the $AdS_7\times S^4$ truncation of \cite{PvN} could
eventually provide  another case, but those results remain to be
analyzed from the  point of view taken here) which has been analyzed
in the above framework.  
In that case, the internal connection
components ${Q_m}^{\!A}{}_{\!B}$ and $P_m^{ABCD}$ do survive the truncation
to the massless modes and are metamorphosed into the $T$-tensor 
describing the couplings of the scalars and the fermions in gauged
supergravity \cite{dewnic3,dewnic2}. The remarkable fact 
is that for gauged $N\!=\!8$ supergravity in four dimensions,
this $T$-tensor does show features related to the ${\rm E}_{7(7)}$
symmetry group, even though ${\rm E}_{7(7)}$ is not a symmetry of the
gauged theory! Namely, the $T$-tensor can be assigned to the $\bf 912$ 
representation of ${\rm E}_{7(7)}$, which branches into SU(8)
representations according to \cite{dewnic4}
\be
{\bf 912} \longrightarrow {\bf 36} \oplus \overline{\bf 36} \oplus
{\bf 420} \oplus \overline{\bf 420}  \,. 
\ee
Similar results have been obtained for three- and five-dimensional
gauged maximal supergravity \cite{GRW,nicsam}. 

\section{MATTER COUPLINGS}
Electric-magnetic duality also plays an important role in
supergravity coupled to vector supermultiplets and in rigidly
supersymmetric gauge theories. Of course, the possibility for
performing these duality transformations is restricted in the presence
of charged fields, because the Lagrangian will then depend explicitly on
vector potentials. Below I will concentrate on the case of $N\!=\!2$
abelian vector multiplets. Before 
doing so, I should briefly draw attention to the case of $N\!=\!4$
supersymmetry. First of all, there is the intriguing conjecture
\cite{MonOli} that there exists a strong-weak coupling duality between
electric charges and magnetic charges in a field theory, or,
equivalently, between conventional local charges and topological
charges. Obviously this is an electric-magnetic duality. It is
believed (see, {\it e.g.} \cite{Osborn}) that this 
duality is realized explicitly for $N\!=\!4$ supersymmetric Yang-Mills
theory. Secondly, I refer to the coupling of $N\!=\!4$ 
vector multiplets to  supergravity, which was worked out in
\cite{deRoo} by coupling on-shell vector multiplets to $N\!=\!4$
conformal supergravity \cite{BDRDW} and imposing electric-magnetic
duality on the equations of motion.  

The actions for $N\!=\!2$ vector multiplets are based on $N\!=2\!$
chiral superspace integrals,
\be
S\propto {\rm Im}\;\Big(\int {\rm d}^4x \;{\rm d}^4\theta\; 
F(W^I)\Big)\,,  \label{chiral}
\ee
where $F$ is is an arbitrary function of reduced chiral multiplets 
$W^I(x,\theta)$. The indices $I,J,\ldots$
will now label the vector multiplets.\footnote{
   In the rigidly supersymmetric case I assume $n$ vector
   multiplets, so that $I,J= 1,\ldots,n$. In the coupling to supergravity
   one has to include the graviphoton. In the superconformal 
   multiplet calculus this implies that one must include one more vector
   multiplet, so that the indices take the values $I,J=0,
   1,\ldots,n$. For supergravity the duality group wil thus be contained
   in ${\rm Sp}(2n+2;{\bf R})$. } 
Such multiplets carry the gauge-covariant degrees of 
freedom of a vector multiplet, consisting of a complex scalar 
$X^I$, a spinor doublet $\Omega^{iI}$, a selfdual field-strength 
$F_{\m\n}^{-I}$ and a triplet of auxiliary fields $Y^I_{ij}$. 
This Lagrangian may coincide with the effective Lagrangian associated  
with some supersymmetric Yang-Mills theory, but for 
our purposes its origin is not directly relevant. To enable 
coupling to supergravity the holomorphic function should be 
homogeneous of second degree \cite{su}.  

As before, I choose abelian gauge groups, so that the field strengths
are subject to electric-magnetic duality 
transformations.  Supersymmetry now dictates how the fields of
a vector multiplet other than
the field strength should transform under electric-magnetic
duality. As a result the structure of the dualities can now 
be captured in terms of the complex scalars $X^I$ and
the holomorphic function $F(X)$. For generic $N=2$ vector supermultiplets
it was discovered \cite{DWVP} that the dualities rotate
the scalar fields $X^I$ and the derivatives  
$F_I$ of the holomorphic function $F(X)$ that encodes the 
Lagrangian, by means of the same ${\rm Sp}(2n;{\bf R})$ matrix that
was introduced before (c.f. \eqn{FGdual}),
\be
\pmatrix{X^{I}\cr  \noalign{\vskip 1mm} F_{I}\cr} \longrightarrow
\pmatrix{\tilde X^I\cr \noalign{\vskip 1mm} \tilde F_I\cr}=
\pmatrix{U&Z\cr \noalign{\vskip 1mm} W&V\cr} 
\pmatrix{X^{I}\cr \noalign{\vskip 1mm} F_I\cr}\,. 
\label{transX}
\ee
Because the matrix belongs again to ${\rm Sp}(2n;{\bf R})$, one can
show that  the new quantities $\tilde F_I$ can be written as  
the derivatives of a new function $\tilde F(\tilde X)$. 
The new but equivalent set of equations of motion are obtained from
the Lagrangian based on $\tilde F(\tilde X)$.   
It is possible to integrate \eqn{transX} and one finds 
\bea
\tilde F(\tilde X) &\,=\,& F(X)-{\textstyle{1\over2}}X^I F_I(X) 
\nonumber \\
&& + {\textstyle{1\over2}} \big(U^{\rm T}W\big)_{IJ}X^I X^J 
\nonumber \\
&&+{\textstyle{1\over2}}
\big(U^{\rm T}V + W^{\rm T}  Z\big)_I{}^J X^IF_J\nonumber \\
&&+{\textstyle{1\over2}} \big(Z^{\rm T}V\big){}^{IJ}F_I F_J \,, 
\label{newfunction}
\eea
up to a constant and terms linear in the $\tilde X^I$. In the coupling to 
supergravity, where the function must be homogeneous of second 
degree, such terms are obviously excluded.\footnote{
   The terms linear in $\tilde X$ in \eqn{newfunction} are 
   associated with constant translations in $\tilde 
   F_I$ in addition to the symplectic rotation shown in 
   \eqn{transX}. Likewise one may introduce constant shifts in
   $\tilde X^I$. Henceforth I ignore these shifts, which are 
   excluded for local supersymmetry. Constant contributions to $F(X)$
   are always irrelevant.} 
The last three terms in \eqn{newfunction} can be written as $\ft12
\tilde X^I \tilde F_I(\tilde X)$, so that the above result is
precisely analogous to the result \eqn{newlagrangian}. Also the  
derivations proceed along similar lines. 

Just as \eqn{newlagrangian}, the above expression \eqn{newfunction} is
not so useful, as it  
requires substituting $\tilde X^I$ in terms of $X^I$, or vice 
versa. I should stress once more that I am not yet dealing with an
invariance, but with an equivalence \cite{CecFerGir}. The functions
$F(X)$ and $\tilde F(\tilde X)$ lead to the same set of field
equations and Bianchi identities. As an example (for more details, see
{\it e.g.} \cite{cerndual}) I note that the following three
functions are equivalent by electric-magnetic duality,
\bea
F_1(X) &=& -{X^1(X^2)^2\over X^0} \,,\nonumber\\
F_2(X) &=&   { X^0( X^2)^2\over X^1} + (X^2)^2 \,, \nonumber \\[1.5mm]
F_3(X) &=& \pm \sqrt{ - X^0 ( X^1)^2  X^2} \nonumber\\
&&+ ( X^1)^2 +  X^1 X^2 +  ( X^2)^2\,.
\eea

When $F$ remains unchanged, $\tilde F(\tilde X) = F(\tilde 
X)$,  the theory is {\it invariant} under the 
corresponding transformations, but again it is hard to verify 
the invariance directly in this form. A more convenient method is to
check whether the substitutions $X^ I\to \tilde X^I$ into the derivatives 
$F_I(X)$ correctly induce the symplectic transformations \cite{DWKLL},
{\it i.e.}, 
\be
F_I(\tilde X) = V_I{}^J F_I(X) + W_{IJ} X^J\,.
\ee
This follows straightforwardly by differentiating \eqn{newfunction}
with respect to $X^I$, using the fact that $F$ remains the same. The
same remark was made in section~1, when discussing general theories
with abelian gauge fields. For
continuous duality symmetries, one derives a condition 
\be
C_{IJ} \,X^IX^J -2 B^I_{\,J}\, F_I X^J
+ D^{IJ}\, F_I F_J =0 \,, \label{consistency3}
\ee
for all $X$. This condition, derived in \cite{DWVP}, is 
the direct analogue of \eqn{consistency2}.

Hence the functions $F(X)$ decompose into equivalence classes. Two different
functions may still describe the same theory, and their equivalence is
effected by electric-magnetic duality. For the subgroup of the 
symplectic group corresponding to an invariance of the equations 
of motion, $F(X)$ remains the same.  
These symplectic reparametrizations were at the basis of the work
initiated in \cite{SW} where exact solutions of low-energy effective
actions for $N=2$ supersymmetric Yang-Mills theories were
obtained. Singularities in these effective actions signal their breakdown 
due to the emergence of massless states corresponding to 
monopoles and dyons.  Although these states are the result of 
nonperturbative dynamcis, they are nevertheless accessible 
because near these singularities one conveniently converts to an 
alternative dual formulation in which local field theory is 
again applicable. In many of the nonperturbative solutions the 
quantities $(X^I,F_J)$ can be 
defined as the periods of a meromorphic differential 
corresponding to a class of hyperelliptic curves (see {\it e.g.}
\cite{KLTY,ArgFar,STY,EY}; see also the talk by Argyres at the
symposium \cite{Argyres}). The same phenomena play a role for vector fields 
coupled to supergravity, for instance, in the context of heterotic string
compactifications \cite{DWKLL,AFGNT}. In the context of type-II string 
compactifications on Calabi-Yau manifolds, the $(X^I,F_J)$ can 
be associated with the periods of the $(3,0)$ form of the 
Calabi-Yau three-fold \cite{Candelas}. A detailed discussion of these
results is outside the scope of this review. However, I should mention
that these systems of vector multiplets give rise to {\it special
geometry}. Again, I refrain from giving further details and refer to
the literature (see, {\it e.g.},
\cite{strominger,Ceresole,DWVPspecial,CRTVP}).  

The result \eqn{newfunction} shows immediately that  $F(X)$ does not
transform as a function under duality, but the combination 
\be
F(X)-{\textstyle{1\over2}}X^I F_I(X) \label{holfunct}
\ee
does, {\it i.e.}, it transforms according to $\tilde f(\tilde X) =
f(X)$. In the case that there are duality {\it invariances}, 
\eqn{holfunct} is an invariant function under the corresponding symplectic
transformations. It should be stressed that, 
although one generically calls quantities such as $F(X)$ 
holomorphic functions, I use the term now in a much more
restricted sense by insisting that certain quantities transform as
functions {\it under symplectic transformations}. Quantities with 
this property are more rare. I stress that physical results
should always be expressible in terms of such functions. The reason is
that an {\it ab initio} calculation of physical quantities will always
yield a function (I assume that the quantity is a scalar). For
instance, suppose that one could directly calculate 
the mass spectrum of solitonic solutions and that the soliton  mass would
depend exclusively on the charges. If this were the case, the function
that expresses the mass would have to be a symplectic function. In
fact, there would be a stronger result: the function should be
{\it invariant} with respect to ${\rm Sp}(2n;{\bf Z})$, simply because the
solutions corresponding to these charges are equivalent up to
electric-magnetic duality. Clearly, this result must be
wrong: there is no symplectic invariant that depends exclusively on
the charges, so the mass formula would be constant! In reality 
the mass will depend also  on the function $F(X)$, which changes under
duality as well. Hence the mass formula will depend on the charges
$(p^I,q_I)$ and on $X^I$ and derivatives of $F(X)$ in such a way that
the expression 
transforms as a function under duality. However, suppose now that we
restrict ourselves to the subgroup of dualities that consitutes an
invariance. In that case, the function $F(X)$ remains the same
and the mass formula must be an {\it invariant} function of 
$X^I$, $p^I$ and $q_I$. These observations have, for instance, been put
to a test for  extremal black holes, where the entropy for BPS black
holes is indeed 
a symplectic function, even when one includes terms quadratic in the
Weyl tensor into the effective low-energy theory \cite{CdWM}. 
At the event horizon the $X^I$ are determined by
the charges according to the so-called fixed-point behaviour noted
in \cite{FKS,Strom,FeKa}. The equations that determine the $X^I$ in
terms of the 
charges are symplectically covariant.\footnote{
   Even for actions of $N\!=\!2$ matter-coupled supergravity with
   terms quadratic in the Riemann curvature the fixed-point behaviour
   has been established \cite{extrbh} . } 
Therefore the entropy is an invariant
function of the electric and magnetic charges\footnote{
  Observe that the {\it central charges} vanish at the horizon because
  supersymmetry is fully restored there. } 
under the duality symmetries. Note that
also the ADM mass formula for the black holes is a symplectic function
and is invariant under the invariant duality subgroup, except that this
function depends both on the charges and on the values of the moduli at
spatial infinity. The above reasoning is not restricted to
$N\!=\!2$ supersymmetric systems and there are many examples in the
literature of duality invariant entropy functions which have been
discussed from a variety of viewpoints, see {\it e.g.}, 
\cite{CvYo,LaWi,CvTs,KaKo,CvHu,BKRSW,CaLuMo,BCDKLM,FeKa,CdWM}.    

\section{DUALITY AND HOLOMORPHY}
I now extend the discussion by considering  supersymmetric Yang-Mills
theories in the presence of a chiral background field \cite{dW96}. To
couple supersymmetric vector multiplets  
to (scalar) chiral background fields is straightforward. One 
simply incorporates additional chiral fields $\Phi$ into the 
function $F$ that appears in the integrand of \eqn{chiral}, 
\be
S\propto {\rm Im}\;\Big(\int {\rm d}^4x \;{\rm d}^4\theta\; 
F(W^I,\Phi)\Big)\,.  \label{chiral2}
\ee
Also the coupling to conformal supergravity is
known \cite{DWLVP}. I draw attention to the fact that the $W^I$ 
are reduced, while the $\Phi$ can be either reduced or general 
chiral fields.  

Let me briefly review situations where such chiral 
backgrounds are relevant. In supersymmetric theories many of the
parameters  (coupling constants, masses) can be regarded as  
background fields that are frozen to constant values (so that 
supersymmetry is left intact). Because these background fields 
correspond to certain representations of supersymmetry, the way 
in which they appear in the theory -- usually both perturbatively 
as well as nonperturbatively -- is restricted by supersymmetry. 
In this way one may derive restrictions on the way in which 
parameters can appear. An example is, for instance, the coupling 
constant and $\theta$-angle of a supersymmetric gauge theory, 
which can be regarded as a chiral field frozen to a complex 
constant $iS= \theta/2\pi+i4\pi/g^2$. 
Supersymmetry now requires that the function $F(X)$ depends on 
$S$, but {\it not} on its complex conjugate. This strategy of 
introducing so-called {\it spurion} fields is not new. In the 
context of supersymmetry it has been used, for instance in 
\cite{shifman,amati,seiberg}, to derive nonrenormalization 
theorems and even exact results. Another option is to not restrict 
the spurion superfields to constant values, in order to introduce 
an explicit breaking of supersymmetry \cite{GG,AGD}. 

The above approach is very natural from the point  
of view of string theory, where the moduli fields, which 
characterize the parameters of the (supersymmetric) low-energy 
physics, reside in supermultiplets. In heterotic $N=2$ 
compactifications the 
background field $S$ introduced above coincides with the complex 
dilaton field, which comprises the dilaton and the axion, and 
belongs to a vector multiplet. The dilaton acts as the 
loop-counting parameter for string perturbation theory. Although 
the full supermultiplet that contains the dilaton is now 
physical, the derivation of nonrenormalization theorems can proceed
in the same way \cite{nilles,DWKLL}. I should stress here that 
when restricting the background to a reduced chiral multiplet, 
one can just treat it as an additional (albeit external) 
vector multiplet. Under these circumstances one may consider 
extensions of the symplectic  
transformations that involve also the background itself. Of 
course, when freezing the background to constant values, one must 
restrict the symplectic transformations accordingly. The above 
strategy is especially useful when dealing with anomalous 
symmetries. By extending anomalous transformations to the 
background fields, the  variation of these fields can compensate 
for the anomaly. The extended non-anomalous symmetry becomes
again anomalous once the background is frozen to a contant 
value. 

Another application of chiral backgrounds, which is
relevant later on in this section, concerns  
the coupling to the Weyl multiplet, which leads to interactions 
of vector multiplets to the square of the Riemann tensor. In this 
case the scalar chiral background  
is not reduced and is proportional to the square of the Weyl 
multiplet. Here the strategy is not to freeze the 
background to a constant, but one is interested in more general 
couplings with conformal supergravity. This application is relevant
for considering actions with higher-derivative interactions involving
the square of the Weyl tensor in supergravity coupled to
matter (see, {\it e.g.} \cite{AGNT1,DCLMR,CdWM}). 

The following discussion is based on the action \eqn{chiral2}. 
A first observation is that, a priori, it is not  
meaningful to restrict the dependence on the background field. For 
instance, one may couple the theory linearly to the background, 
so that $\cal N$ will depend at most linearly on the background 
field. However, after a symplectic transformation, $\cal N$ will 
generically have a nonlinear dependence on the background, as 
follows from \eqn{nchange}. Therefore 
the only meaningful approach is to start from functions $F$ which 
depend both on the superfields $W^I$ and on the background 
field $\Phi$ in a way that is a priori unrestricted. Then one can 
proceed exactly as before and examine the symplectic equivalence
classes in the presence of the background. The transformation rules, 
however, will also depend on the background fields, because $X^I$
transforms into $F_I$ and the latter is background dependent. This
does not affect the derivation, although there are a number of new 
features. 

The analysis of electric-magnetic duality in the presence of the chiral
background proceeds in the same way as before (for more details, see
\cite{dW96}). Denoting the lowest 
$\theta$-component of $\Phi$ by $\hat A$, one is dealing with functions 
$F(X,\hat A)$. One starts from the same transformation rule \eqn{transX} 
and obtains the same expression \eqn{newfunction} for the new function 
after the symplectic transformation. But now the relation between $X$ and 
$\tilde X$ involves $\hat A$. In the same notation as used previously I
note the relation\footnote{%
   In supergravity the expression for $\cal N$ changes when 
   integrating out the auxiliary tensor field. This is not possible in
   theories with higher-derivative couplings, so I refrain from doing
   this. Therefore I must insist that the function 
   $F$ exists. After integrating out the auxiliary tensor it is 
   possible to reformulate the theory in such a way that the 
   function $F$ no longer needs to exist, as long as the periods 
   $(X^I,F_J)$ can  be written down consistently \cite{Ceresole}.}%
\be
{\cal N}_{IJ} =\bar F_{IJ}\,,
\ee
It is convenient to introduce the following definitions, 
\bea
{\partial\tilde X^I\over\partial X^J}\,&\equiv& \,{\cal 
S}^I{}_{\!J}(X,\hat A)
= U^I{}_{\!J} +Z^{IK}\,F_{KJ} \,, \nonumber\\
{\cal Z}^{IJ}\,&\equiv& \,[{\cal S}^{-1}]^I{}_K\, Z^{KJ}\,, \nonumber\\ 
N_{IJ}\,&\equiv&\, 2 \,{\rm Im} \,F_{IJ}\,, \qquad N^{IJ}\equiv 
\big[N^{-1}\big]^{IJ}\,,
\eea
which all depend on the background.
The quantity ${\cal Z}^{IJ}$ is symmetric in $I$ and $J$, because 
$Z\,U^{\rm T}$ is a symmetric matrix as a consequence of the fact 
that $U$ and $Z$ are submatrices of the symplectic matrix. 

On the quantities ${\cal N}_{IJ}$ and ${\cal O}^\pm_{\m\n I}$ that were
introduced in \eqn{FF+FO}, the symplectic reparametrizations act
according to (\ref{nchange},\ref{ochange}), which, in the above
notation, read as follows, 
\bea
\tilde{\cal N}_{IJ} \,&=&\, (V_I{}^K {\cal N}_{KL}+ W_{IL} )\,
[\bar{\cal S}^{-1}]^L{}_J \,,\nonumber \\  
\tilde{\cal O}_{\mu\nu I}^+ \,&=&\, {\cal O}_{\mu\nu J}^+  \,
[\bar{\cal S}^{-1}]^J{}_I \,, \label{NOchange}
\eea
irrespective of the presence of the background. Also the following result,
\bea
&&\tilde F(\tilde X,\hat A)-{\textstyle{1\over2}}\tilde X^I \tilde 
F_I(\tilde X,\hat A) =\nonumber \\
&&\qquad \qquad F(X,\hat A)-{\textstyle{1\over2}} X^I 
F_I(X,\hat A) \,,  \label{holfunct2}
\eea
still holds, so that there is a holomorphic function that 
transforms as a function under symplectic transformations. In the 
coupling to supergravity this result is still relevant, provided  
the background field $\hat A$ has a nonzero scaling weight. 
Other results which hold irrespective of the background, are 
\bea
\tilde N_{IJ} &=& N_{KL}\, \big[\bar{\cal 
S}^{-1}\big]^K{}_{\!I}\,  \big[{\cal S}^{-1}\big]^L{}_{\!J}\,, 
\nonumber\\ 
\tilde N^{IJ} &=& N^{KL}\, \bar{\cal 
S}^I{}_{\!K}\,{\cal S}^J{}_{\!L}\,,\nonumber \\  
\tilde F_{IJK} &=& F_{MNP}\, \big[{\cal S}^{-1}\big]^M{}_{\!I} \,
    \big[{\cal S}^{-1}\big]^N{}_{\!J}\, \big[{\cal 
S}^{-1}\big]^P{}_{\!K} \,, 
\eea
where all quantities depend on both the fields $X^I$ and $\hat A$. The
symmetry in $I$ and $J$ of the first two quantities is preserved  
owing to the symplectic nature of the transformation. 
Results that specifically refer to the background are obtained 
by taking derivatives of $\tilde F$ (cf. \eqn{newfunction}), 
keeping $\tilde X^I$ fixed in partial differentiations of  
$\tilde F$ with respect to $\hat A$, and/or using already known 
transformations. In this way one obtains, for instance,
\bea
\tilde F_A(\tilde X,\hat A)&=&  F_A(X,\hat A)\,, \nonumber\\
\tilde F_{AI} &=& F_{AJ}\,[{\cal S}^{-1}]^J{}_I \,,\nonumber\\
\tilde F_I-\tilde F_{IJ}\tilde X^J  &=& [ F_J- F_{JK} X^K]\,
[{\cal S}^{-1}]^J{}_I\,, \nonumber\\
\tilde F_I-\tilde{\bar F}_{IJ}\tilde X^J  &=& [F_J-{\bar F}_{JK} 
X^K]\,[\bar{\cal S}^{-1}]^J{}_I \,, \nonumber\\
\tilde F_{AA}(\tilde X,\hat A)&=& F_{AA}(X,\hat A) \nonumber\\
&&-F_{AI}(X,\hat 
A)\,F_{AJ}(X,\hat A)\, {\cal Z}^{IJ}\,. \label{vgl1}
\eea
Observe that we
have now identified two functions that are both symplectic {\it and}
holomorphic. No other functions of this type are known. In the
coupling to supergravity, these functions are 
actually related, provided that $\hat A$ has nonzero scaling weight,
because in that case $F(X,\hat A)$ must be homogeneous. 
Note that the generalized $N\!=\!2$ black hole entropy formula \cite{CdWM}
turns out to consist of two terms, $\bar X^IF_I - \bar F_IX^I$ and
$F_{\hat A}$, which are both symplectic functions (we suppress details
related to the supergravity coupling). When dealing with a duality
symmetry, both functions are invariant. The result that $F_A$ is
invariant is somewhat analogous to a result of \cite{GZ}, according
to which the derivative of the Lagrangian with respect to a duality
invariant parameter is invariant. 

However, higher than first derivatives of $F$ with respect to $\hat A$ 
do {\it not} transform as functions under symplectic transformations. 
This means that the coefficient functions in an expansion such as 
\eqn{CYW2} do not transform as symplectic functions, with the 
exception of $F_A$. This result may be somewhat disturbing especially
when considering  
symplectic transformations that constitute an invariance. In that 
situation one  has $\tilde F(\tilde X,\hat A) = F(\tilde X,\hat A)$. 
In spite of that, this does not imply that the coefficient 
functions (i.e.  multiple derivatives with respect to the 
background) are invariant {\it functions} under the corresponding 
tranformations. This is only the case for the first one 
corresponding to $F_A$.  

One may wonder whether there exist modifications of the 
multiple-$\hat A$ derivatives of $F$ that do transform as 
functions under symplectic transformations. Such functions should 
be expected to arise when evaluating the coefficient functions 
directly on the basis of 
some underlying theory, such as string theory. Modifications seem
possible in view of the fact that the combination 
$$
F_{AA} + i N^{IJ} F_{AI} F_{AJ} 
$$
does indeed transform as a function under symplectic
transformations. Likewise, one may verify by explicit calculation
that there is a generalization of the third  derivative, 
\bea
F_{AAA} &&+ 3 i N^{IJ}F_{AAI}F_{AJ} \nonumber\\
&&  - 3 N^{IK} \,N^{JL}\,F_{AIJ}F_{AK}F_{AL}\nonumber\\  
&& - i N^{IL}\,N^{JM}\, N^{KN} \,F_{IJK} 
F_{AL}F_{AM}F_{AN}\,,\nonumber 
\eea
which also transforms as a symplectic function. 

It turns out that these functions can be generated 
systematically. Assume that $G(X,\hat A)$ transforms as a 
function under symplectic transformations. Then one readily 
proves that also ${\cal D}G(X,\hat A)$ transforms as a symplectic 
function, where\footnote{%
   I note that ${\cal D}$ and $N^{IJ}\pa_J$ commute. } %
\be 
{\cal D}\,\equiv\, {\pa\over \pa \hat A} +
iF_{AI}N^{IJ}{\pa\over \pa X^J} \,. 
\ee
Consequently one can write down a hierarchy of functions which are
modifications of  multiple derivatives $F_{A\cdots A}$,
\be
F^{(n)} (X,\hat A) \,\equiv \, {1\over n!}{\cal D}^{n-1} F_A(X,
\hat A)\, ,\label{Fn}
\ee
where I included a normalization factor. 
All the $F^{(n)}$ transform as functions under symplectic 
functions. However, except for the first one, they  
are {\it not} holomorphic. The lack of holomorphy is governed 
by the following equation ($n>1$),
\be
{\pa F^{(n)} \over \pa \bar X^I}= \ft12 \bar F_I{}^{JK} 
\sum_{r=1}^{n-1}\; {\pa F^{(r)} \over \pa  X^J}\,{\pa F^{(n-r)} 
\over \pa  X^K}\,, \label{anomalyeq2}
\ee
where $\bar F_I{}^{JK}= \bar F_{ILM}\,N^{LJ}N^{MK}$. 

Interestingly enough, this equation is reminiscent of the holomorphic
anomaly equation of \cite{BCOV}. To explain the relation let us
discuss the coupling to supergravity and 
associate the chiral background field $\Phi$ with the square 
of the Weyl multiplet $\cal W$, thus obtaining higher-derivative
couplings of vector multiplets with conformal supergravity. 
The square of the Weyl multiplet constitutes a scalar chiral 
field of scaling weight 2. Its lowest component is equal to 
$(\varepsilon_{ij} T^{ij}_{ab})^2$, where $T^{ij}_{ab}$ is sometimes
(incorrectly) referred to as the graviphoton field strength; the
highest-$\theta$ component contains the square of the selfdual
components of the Riemann tensor. I refer to \cite{BDRDW} for details. 
Now assume that the function $F$ can be expanded as a power series,
\be
F(X, {\cal W}^2) = \sum_{g=0}^\infty \;F^{(g)}(X)\, \big({\cal 
W}^2\big)^g \,.\label{CYW2}
\ee
Because it must be homogeneous of second 
degree with scaling weights of $X$ and $\cal W$ that are both equal 
to unity, the coefficient functions $F^{(g)}(X)$ are homogeneous 
of degree $2(1-g)$. 
  
In supergravity the $X^I$ are not independent scalar fields, but 
are defined projectively; in more mathematical terms they can be 
regarded as sections of a complex line bundle. 
These sections can be expressed holomorphically in terms of 
independent complex fields $z^A$, which describe the physical 
scalars of the vector multiplets. The original quantities $X^I$ and 
the holomorphic sections $X^I(z)$ differ by a factor $m_{\rm 
P}\exp(K/2)$, where $K$ is the K\"ahler potential\footnote{%
   In terms of the holomorphic sections the K\"ahler potential 
   takes the form 
   $$
   K(z,\bar z)= -\log\big(i \bar X^I(\bar z)\,F_I(X(z)) -i X^I(z)\,
   \bar F_I(\bar X(\bar z))\big) \,.
   $$ 
   The K\"ahler metric is defined as $g_{A\bar B}= 
   \pa_A\pa_{\bar B} K(z,\bar z)$. Under projective 
   transformations of the 
   holomorphic sections, $X^I \to \exp (f(z)) \,X^I$, the K\"ahler 
   potential transforms by a K\"ahler transformation, so that the 
   metric remains invariant.} %
and $m_{\rm P}$ is the Planck mass. 
In view of the projective nature of the $X^I$, there 
is thus always one more physical vector field than there are physical 
scalars. The extra vector corresponds to the graviphoton.  
The Lagrangian encoded by \eqn{CYW2} gives rise to 
terms proportional to the square of the Riemann tensor times 
$(\varepsilon_{ij} T^{ij}_{ab})^{2(g-1)}$. After extracting the 
scale factor $m_{\rm P}\,\exp(K/2)$, the coefficient  
functions $F^{(g)}(X)$ give rise to holomorphic functions ${\cal 
F}^{(g)}$ of the $z$ (or rather sections of a line bundle).    

Now consider the case where the function \eqn{CYW2} encodes the
$N\!=\!2$ supersymmetric effective low-energy field theory
corresponding to a type-II  
string compactification on a Calabi-Yau manifold. 
In that case one can show that the ${\cal F}^{(g)}$ represent 
$g$-loop contributions in string perturbation theory.\footnote{
  In $N=2$ compactifications of the heterotic string the counting 
  of string loops runs differently. } 
The coefficient functions can be determined in string theory from
certain type-II string amplitudes  \cite{AGNT1} and indeed arise in
the appropriate orders in string perturbation theory. An interesting 
feature is that the ${\cal F}^{(g)}$ can be 
identified with the topological partition function of a twisted 
nonlinear sigma model on a Calabi-Yau target space, defined on a 
two-dimensional base space equal to a genus-$g$ Riemann surface. 
The partition function is obtained by integrating 
appropriately over all these Riemann surfaces 
\cite{BCOV}. However, the partition functions 
${\cal F}^{(g)}$ do not depend holomorphically on the Calabi-Yau 
moduli. They  exhibit a 
holomorphic anomaly due to the propagation of massless states, or 
equivalently, due to certain contributions from the boundary of the 
moduli space ${\cal M}_g$ associated with the genus-$g$ Riemann 
surfaces. 

The holomorphic anomaly of \cite{BCOV} receives contributions from two
terms. One is precisely as in \eqn{anomalyeq2} and arises from  
pinchings that separate the Riemann surface into two 
disconnected surfaces. A second term, absent in
\eqn{anomalyeq2}, corresponds to Riemann surfaces where a closed loop
is pinched such that the genus is lowered by one unit. 
We should stress that \eqn{anomalyeq2} was obtained in a 
very general context and applies to both rigid and local $N\!=\!2$ 
supersymmetry. The conclusion is that part of the holomorphic anomaly
can thus be viewed as  
arising from a conflict between the requirements of holomorphy 
and of a proper (covariant) behaviour under symplectic
transformations. The nonholomorphic modifications exhibited above can
be regarded as (part of) the threshold corrections that arise due to
the propagation of massless states \cite{DKL}. 

\vspace{5mm}              
\par                      
\noindent{\bf Acknowledgements}\vspace{0.3cm} 
\par                      
I thank Marc Grisaru, Hermann Nicolai and Toine Van Proeyen for making
valuable suggestions concerning the text.



\end{document}